\documentclass{INTERSPEECH2023}

\interspeechcameraready

\title{Memory-augmented conformer for improved end-to-end long-form ASR\thanks{This work was supported by Portuguese national funds through Fundação para a Ciência e a Tecnologia (FCT), with references UIDB/50021/2020 and 2022/12328/BD, as well as by the Portuguese Recovery and Resilience Plan (RRP) through project C644865762-00000008 (Accelerat.AI)}}
\name{Carlos Carvalho$^{1,2}$, Alberto Abad$^{1,2}$}
\address{
  $^1$INESC-ID, Lisbon, Portugal\\
  $^2$Instituto Superior T\'{e}cnico, Universidade de Lisboa, Portugal
  }
\email{carlos.mf.carvalho@inesc-id.pt, alberto.abad@inesc-id.pt}

\begin{document}

\maketitle

\begin{abstract}
Conformers have recently been proposed as a promising modelling approach for automatic speech recognition (ASR), outperforming recurrent neural network-based approaches and transformers. Nevertheless, in general, the performance of these end-to-end models, especially attention-based models, is particularly degraded in the case of long utterances.\ To address this limitation, we propose adding a fully-differentiable memory-augmented neural network between the encoder and decoder of a conformer.\ This external memory can enrich the generalization for longer utterances since it allows the system to store and retrieve more information recurrently.\ Notably, we explore the neural Turing machine (NTM) that results in our proposed Conformer-NTM model architecture for ASR. Experimental results using Librispeech train-clean-100 and train-960 sets show that the proposed system outperforms the baseline conformer without memory for long utterances.
\end{abstract}

\section{Introduction}

Traditional speech recognition systems rely on sophisticated and individual components, including acoustic, pronunciation and language models (LMs) \cite{hmms}.\ In contrast, end-to-end (E2E) speech recognition systems rely on a single deep neural network that learns to map an input sequence of features or raw audio to the corresponding labels; usually, characters or sub-words \cite{pmlr-v32-graves14, 8706675}.\ Because of this simplicity, and in some situations, superior performance over traditional systems, E2E systems have become a favoured procedure for automatic speech recognition (ASR) \cite{ 7472621, chiu2018state}.\ Some widely used E2E approaches are based on connectionist temporal classification (CTC) \cite{graves2006connectionist}, recurrent neural network transducers (RNN-T) \cite{graves2012sequence} and attention-based encoder-decoders (AEDs) \cite{7472621}.

The transformer \cite{vaswani2017attention} architecture is an AED-based system that uses self-attention to capture long-range interactions.\ Nevertheless, a transformer has more difficulty extracting fine-grained local feature patterns than convolution neural networks (CNNs) \cite{49414}.
For this reason, conformers \cite{49414} have been proposed as an approach for E2E ASR, which outperform RNN-based approaches and transformers since they can model the global and local dependencies of an audio sequence by combining CNNs with transformers.
Nonetheless, E2E ASR methods, particularly AED-based procedures, are known to degrade performance on long utterances when trained on short utterances \cite{narayanan2019recognizing, chiu2019comparison}.\  Besides, long-form transcription is crucial for creating continuous transcriptions of real-world scenarios, like lectures, meetings, and video captions (e.g. YouTube).

The problem of long-form speech has been addressed in some previous works by simulating training on longer utterances \cite{hori2020transformer, hori2021conformer}. 
For example, \cite{hori2020transformer} proposed a method where the transformer or conformer accepts multiple consecutive utterances simultaneously and predicts an output for the last utterance only. This procedure is repeated with a sliding window using one-utterance shifts to recognise the whole recording. Another solution is to segment the audio in advance using a separate voice activity detector (VAD) based approach \cite{bain2023whisperx}, 
or an E2E model that learns to predict segment boundaries \cite{51460}.
The E2E segmenter proposed in \cite{51460} relies on human-created heuristics to insert end-of-segment tokens in utterances at training time so that the model can learn to predict those tokens. 

Only few works try to improve the generalisation of E2E ASR systems to long speech without requiring some pre-processing stage or changing how the model trains and decodes compared to traditional E2E ASR. 
For instance, \cite{10045036} proposes the replacement of self-attention with fast attention, which improves the model generalisation ability for long utterances.

In contrast to the works mentioned above, 
we hypothesise whether adding a memory-augmented neural network (MANN) in between the encoder and decoder module -- like a neural Turing machine (NTM) \cite{graves2014neural} -- may be a convenient method to enrich the learning capacity of a conformer, contributing to increase the network generalisation for longer utterances without the need for any \emph{ad hoc} pre-processing or optimisation in training or decoding.\ 
Thus, NTMs have demonstrated superior performance over long short-term
memory cells (LSTMs) in several learning tasks.\ Moreover, to our knowledge, few  works have investigated the use of MANNs for the E2E ASR task.\ In particular, NTM has been used to perform unsupervised speaker adaptation in \cite{sari2020unsupervised} by storing i-vectors \cite{dehak2010front} and then reading from the memory to combine the resulting read vector with the hidden vectors of the encoder of the listen, attend and spell (LAS) architecture \cite{7472621}.\ However, in that work, the write operation of the NTM was not explored, therefore not taking advantage of the full potential of the external memory. 

In this work, we propose incorporating a MANN based on NTM to improve the generalisation of the offline E2E ASR system to long sentences. We refer to this newly proposed ASR architecture as Conformer-NTM\footnote{\url{https://github.com/Miamoto/Conformer-NTM.git}}. This proposed model and the state-of-the-art (SOTA) conformer baseline (without memory) are trained on Librispeech \cite{panayotov2015librispeech} 100 hours clean and 960 hours.\ Then, we use the test clean and other partitions to evaluate the overall performance of all models. We follow this with an ablation study, testing the models with different utterance lengths (long and very-long).  
Our results show that the E2E system can generalise better with the external memory for longer utterances with the Conformer-NTM model.\ Notice that while the focus of this work is on offline ASR settings, the proposed MANN is also expected to complement streaming ASR approaches that address the long-form ASR problem \cite{narayanan2019recognizing, wu2020streaming, tsunoo2019transformer}.

The rest of the paper is organised as follows.\ Section 2 summarises the MANN system.\ Section 3 introduces the proposed approach.\ In Section 4, we describe the experimental evaluation and obtained results, and in Section 5, we provide some concluding remarks and possible directions for future work.

\section{Memory-augmented neural networks}

MANNs refer to a class of neural networks equipped with external memory that can help improve the learning capability of the neural network \cite{graves2014neural,graves2016hybrid, pmlr-v48-santoro16}. Examples of MANNs are the NTM \cite{graves2014neural}, described below and the differentiable neural computer (DNC) \cite{graves2016hybrid}.


\subsection{NTM model}
\label{ssec:ntm}

NTMs \cite{graves2014neural} can read and write arbitrary content to memory cells without supervision by using a soft-attention mechanism. Moreover, they are fully differentiable, which makes it possible to train them in an E2E fashion.

The overall architecture of the NTM is composed of a controller network, e.g., a deep neural network or RNN, that receives inputs, reads vectors and emits outputs in response.\ 
The controller reads and writes from an external memory matrix via a set of parallel read and write heads.\ Additionally, the controller network emits a set of vectors and values for each individual read (e.g., $ \mathbf{k}_{t}$,  $\beta_{t}$, $g_{t}$, $\mathbf{s}_{t}$  and $\gamma_{t}$) and write head (e.g., $\mathbf{a}_{t}, \mathbf{e}_{t},
\mathbf{k}_{t}$,  $\beta_{t}$, $g_{t}$, $\mathbf{s}_{t}$ and $\gamma_{t}$), detailed below, to help in the reading and writing operations. 

The memory is a matrix $\mathbf{M} \in R^{N \times W}$, where $N$ is the number of memory locations (rows) and $W$ is the vector size of each memory location (columns). The read operation at time step $t$ is the weighted average sum of all memory locations, i.\,e.,

\begin{equation}
    \mathbf{r}_{t} = \sum_{i} \mathbf{w}_{t}^{read}(i) \mathbf{M}_{t}(i) ,
    \label{eq1}
\end{equation} where $\mathbf{w}_{t}^{read}(i)$ is the weight associated to row $i$, and $\mathbf{M}_{t}(i)$ is the memory vector from row $i$. Also, the sum of all weights adds up to one. The write operation at time step $t$ contains two main steps. The first step is an erase phase, i.\,e.,

\begin{equation}
    \mathbf{M}_{t}(i)^{'} = \mathbf{M}_{t-1}(i)[\mathbf{1} - \mathbf{w}_{t}^{write}(i) \mathbf{e}_{t}] ,
    \label{eq2}
\end{equation} where $\mathbf{1}$ is a vector of ones and $\mathbf{e}_{t} \in R^W$ is an erase vector. Last, the second step is an add phase:

\begin{equation}
    \mathbf{M}_{t}(i) = \mathbf{M}_{t}(i)^{'} + \mathbf{w}_{t}^{write}(i)\mathbf{a}_{t} ,
    \label{eq3}
\end{equation} where $\mathbf{a}_{t} \in R^W$ is an add vector. 

The weights mentioned above for reading and writing are computed using the same addressing mechanism in parallel for each of the two heads. For this reason, we explain the  addressing process in general terms. Overall, the addressing mechanism combines two main addressing mechanisms:\ \textit{content-based addressing} and \textit{location-based addressing}.\ The first step to computing the weights is to measure the similarity between $\mathbf{k}_{t}$, outputted by the controller, and each entry of the memory, $\mathbf{M}_{t}(i)$, by using cosine similarity:

\begin{equation}
    K[\mathbf{u},\mathbf{v}] = \frac{\mathbf{u} \cdot \mathbf{v}}{\left\Vert \mathbf{u}\right\Vert_{2} \left\Vert \mathbf{v} \right\Vert_{2}}.
    \label{eq4}
\end{equation}

By applying cosine similarity and softmax over the rows, $\mathbf{M}_{t}(i)$, the computation of the weights using the content-based addressing mechanism is obtained following:
\begin{equation}
    \mathbf{w}^{c}_{t}(i) =  softmax(\beta_{t}K[\mathbf{k}_{t}, \mathbf{M}_{t}])_{i},
     \label{eq5}
\end{equation} where $\beta_{t}$, outputted by the controller, is a positive scalar parameter that determines how concentrated the content weight vector should be.\ The following three steps are focused on location-based addressing.\ The second stage creates $\mathbf{w}^{g}_{t}$ by interpolating $\mathbf{w}^{c}_{t}$ with the weight vector from last time step, $\mathbf{w}_{t-1}$, using $g_{t} \in (0,1)$, also outputted by the controller. This interpolation operation allows the system to learn when to use or ignore content-based addressing. Next, for the focus of the weights to be shifted to different rows, the controller emits a shift weighting vector, $\mathbf{s}_{t}$, that defines a normalised distribution over the allowed integer shifts. Each element in this vector gives the degree to which different integer shifts can occur. The actual shift occurs with a circular convolution:

\begin{equation}
    \mathbf{w}^{*}_{t}(i) = \sum_{j=0}^{N-1} \mathbf{w}^{g}_{t}(j) \mathbf{s}_{t}(i-j).
    \label{eq6}
\end{equation}

Next, there is a sharpening parameter $\gamma_{t} \geq 1$, outputted by the controller, which controls the sharpness of the vector weights:

\begin{equation}
    \mathbf{w}_{t}(i) =  softmax(\mathbf{w}_{t}^{*\gamma_{t}})_{i}.
    \label{eq7}
\end{equation}

Finally, we have a weight vector, $\mathbf{w}_{t}$, that determines where to read from and write to in memory, depending on the specific head.

\begin{figure*}[ht!]
  \centering
  \includegraphics[width=0.9\textwidth]{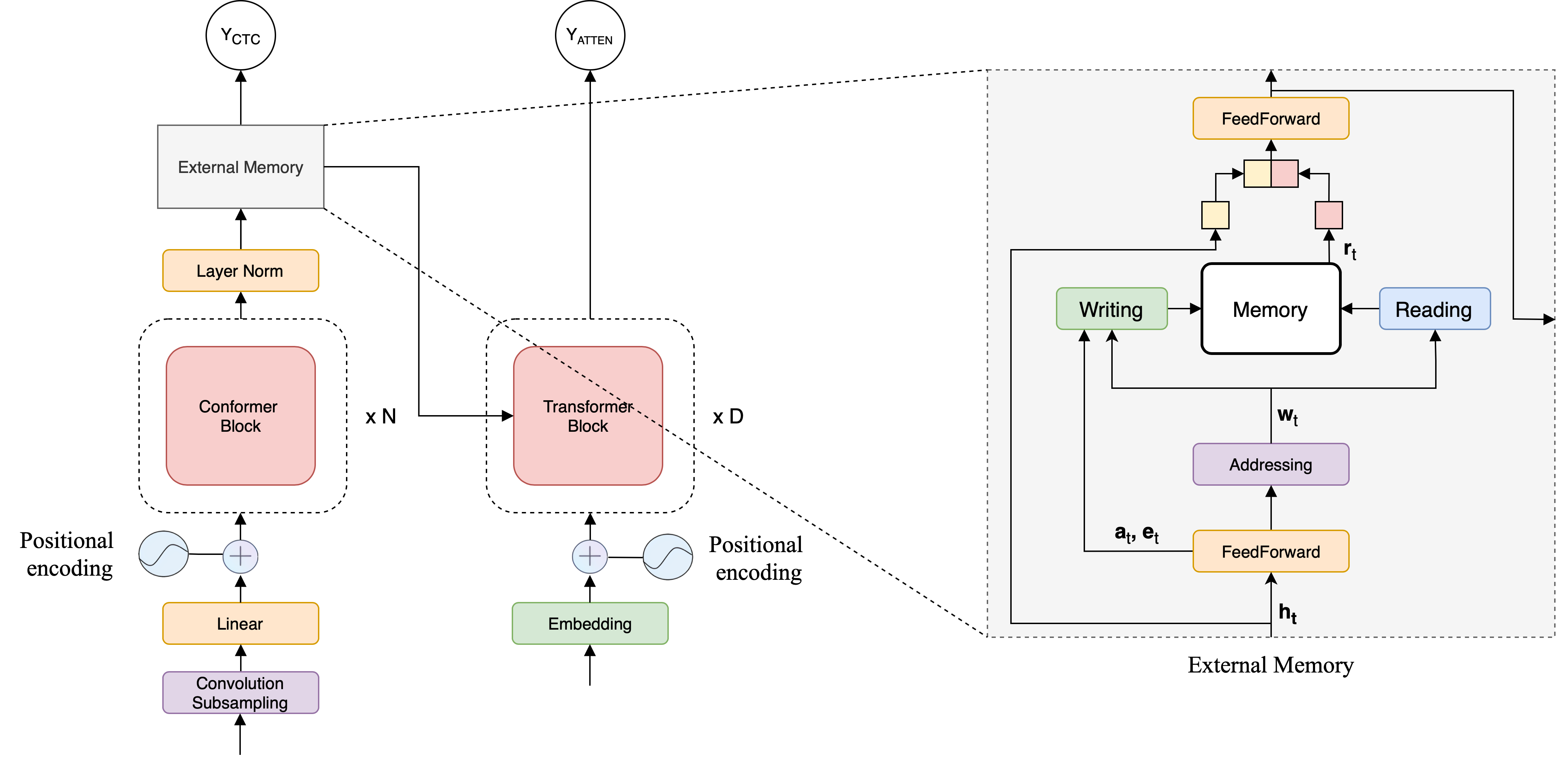}
  \caption{Conformer E2E ASR system with an external fully-differentiable memory.}
  \label{fig:main_model}
\end{figure*}

\section{E2E ASR with a MANN}
\label{sec:pagestyle}
The main structure of our memory-based E2E ASR network is depicted in Figure \ref{fig:main_model}.\ In this architecture we add the proposed external memory in between the encoder and decoder of the Conformer system. 

At first, the input goes through the encoder module, which contains: a convolution subsampling module, a linear projection layer, a relative positional encoding module, N conformer blocks and, at last, a layer norm.\ Then, the output of the layer norm, $\mathbf{h}$, goes trough the external memory system. For each time step $t$, the vector $\mathbf{h_t}$ will be transformed with a feedforward layer for each write and read head so that it is possible to obtain all vectors and values required to help in the reading and writing operations, as mentioned with more detail in Section ~\ref{ssec:ntm}. Following, the attention weights for the read and write head are computed via the addressing mechanism  as described in Section ~\ref{ssec:ntm}.       


Writing and reading occur at each time step, allowing the system to memorise long-term acoustic dependencies recurrently.\ After the read operation, the read vector is concatenated with the encoder output $\mathbf{h}_{t}$.\ Then, this concatenated vector is transformed into a new output vector with the same size of $\mathbf{h}_{t}$ by going through a feedforward layer. Next, the sequence of outputs of the memory goes into the decoder.\ In addition, the decoder also receives the transcription outputs shifted right as input, which go through an embedding layer and a positional encoding module.
The decoder contains D transformer blocks.    

At last, the E2E ASR system combined with the external memory learns in a fully-differentiable way by utilising the joint CTC-attention objective \cite{kim2017joint}.

\section{Experiments}
\label{sec:typestyle}
\vspace{+0.15cm}
\subsection{Data and Experimental Setup}
\label{subsec:setup}
\vspace{-0.15cm}

\begin{table}[th]
\setlength{\tabcolsep}{3pt}

\caption{Mean, minimum (Min.) and maximum (Max.) duration in seconds of the training sets, the test clean, test other, and the subsets long - 100 and very long - 100.}
  \label{tab:statistics}
  \centering
 \begin{tabular}{cc|ccc}
\toprule
\bf Set/Subset & \bf Category& Mean& Min.&Max.        \\
\bottomrule
\bottomrule
train-clean-100 &full set& 12.69& 1.41    & 24.53 \\
train-960 &full set& 12.30& 0.83    & 29.74 \\
test clean &full set& 7.42 & 1.29    & 34.96 \\ 
test other &full set& 6.54 & 1.25    & 34.51 \\ 

\hline
test clean &long - 100& 23.59 & 19.86   & 34.96 \\

test other &long - 100&21.11&  17.13  & 34.51  \\

\hline
concat-clean &very long - 100& 33.41 & 25.39   & 68.14 \\

concat-other &very long - 100&30.76&  21.77  & 68.98  \\

\bottomrule
  \end{tabular}
\end{table}

Our experiments use the Librispeech corpus exclusively.\ We train with the train-clean-100 subset from Librispeech, with 28539 utterances and 585 speakers, and the train-960 set containing 281241 utterances and 5466 speakers.\ We use the dev clean and dev other, with 5567 utterances and 188 speakers.
Finally, we report word-error-rate (WER) and character-error-rate (CER) results for both test clean and test other, with 2620/2939 utterances and 87/90 speakers, respectively.

We also perform an ablation study varying the distribution of the test data conditions, especially considering the utterance length (long and very long).\ For the long setting, we created subsets from the test clean and the test other containing only the longest 100 utterances with the script subset\_data\_dir.sh, from Kaldi \cite{Povey_ASRU2011}.\ For the very long set, we used the original time segmentation information of the Librispeech corpus and concatenated the continuous segments present in the original test clean and the test other sets. Then, we selected the 100 longest concatenated segments using the same Kaldi script mentioned above, resulting into two new subsets: concat-clean and concat-other. Information about the average, minimum and maximum length of the utterances of these subsets are present in Table \ref{tab:statistics}.    

ESPnet2 \cite{watanabe2018espnet} is the toolkit we use to implement and investigate our proposed methods.\ Also, we use a default ESPnet conformer recipe from Librispeech to run all our setups.\ The conformer baseline and the Conformer-NTM models were trained on 2 NVIDIA GeForce RTX 3080. For the 100 hours setup, all models were trained for 80 epochs, while for the 960 hours setting, the models were trained for 50 epochs. In both settings, an average of the ten best checkpoints in the dev set was used as the final model.\ Notice that the Conformer-NTM requires longer training time when compared to the baseline.

The conformer baseline model extracts 80-dimensional FBANK acoustic features on the fly, followed by SpecAugment \cite{Park_2019} and global mean and variance normalization.\ The raw input data has speed perturbation factors of 0.9, 1.0 and 1.1.\ Additionally, the model uses as targets 5000-byte pair encoding (BPE) \cite{sennrich-etal-2016-neural} unigram units learned from the training data.

The encoder is composed of one Conv2D module followed by 12 conformer blocks.\ The Conv2D includes two 2D-CNN layers with 256 channels, a kernel size of 3 x 3, a stride of size two and a ReLU activation.\ 
The decoder is composed of six transformer blocks.\ Both the encoder and decoder have four attention heads of dimension 256.\ The hidden dimension of the feedforward layer for the encoder and decoder is 256, while the output dimensions are 1024 and 2048, respectively.\ Also, the Adam optimizer was used with a learning rate of 0.002, a weight decay of 0.000001 and 15000 warmup steps.\ The number of batch bins for both training setups (train-clean-100 and train-960 hours) was 16 million. Also, the CTC weight was set to 0.3 for training and testing time. 

\begin{table*}[t]
\setlength{\tabcolsep}{4pt}
\caption{(WERs [\%] (CERs [\%]) on the E2E ASR proposed models -- trained with train-clean-100 and train-960 sets from Librispeech -- for the test other and clean sets (full set, long - 100 and very long - 100).}
  \label{tab:wer-cer-results}
  \centering
 \begin{tabular}{cc|cc|cc|cc}
\toprule &\multicolumn{1}{c}{}
 &\multicolumn{2}{c}{Full Set}
&
\multicolumn{2}{c}{Long - 100} &
\multicolumn{2}{c}{Very Long - 100} \\\cmidrule(r){3-4}\cmidrule(l){5-6}\cmidrule(l){7-8}
\bf Model& \bf LM &test clean& test other    & test clean &test other & concat-clean &concat-other \\
\toprule
Conformer-100h &-&6.5 (2.6)&  17.4 (8.5)  & 7.5 (3.4)&  16.6 (8.0) & 12.2 (6.5) & 24.5 (14.3)      \\
\hline
Conformer-NTM-100h &-&6.4 (2.4)&  17.2 (8.4)  & 6.8 (2.6)&  15.8 (7.3) & 9.2 (4.1)  &       21.7 (12.2)      \\
\hline
Conformer-960h &-& 2.8 (1.0)&  6.7 (2.9)  &  3.3 (1.2)&  6.5 (2.7) & 10.0 (5.8) & 15.3 (9.8)    \\
\hline
Conformer-NTM-960h &-& 2.7 (0.9)&  6.8 (2.8)  &  3.0 (0.9)&  5.8 (2.1) &  4.3 (2.0) &  11.6 (6.9)    \\
\toprule
\toprule
Conformer-100h &6-gram&  5.3 (2.3)&  14.7 (7.6)    & 6.3 (3.0)& 14.2 (7.1) & 10.0 (5.9) &  21.2 (12.9) \\
\hline
Conformer-NTM-100h &6-gram&  5.2 (2.1)&  14.5 (7.4) & 5.5 (2.3)& 13.6 (6.5) & 7.4 (3.7) &  18.9 (11.2) \\
\hline
Conformer-960h &6-gram& 2.4 (0.8) &   5.8 (2.5)   & 2.7 (0.9)& 5.3 (2.4) &  9.3 (5.5) & 14.7 (9.7)  \\
\hline
Conformer-NTM-960h &6-gram& 2.4 (0.8) & 5.9 (2.5) & 2.7 (0.8)& 4.8 (1.9) &  3.9 (1.8) & 10.8 (6.7)  \\
\bottomrule
  \end{tabular}
\end{table*}

Regarding the external memories, we first experimented with a different number of rows (128, 256, 512) and columns (5,8,10,40) for the NTM using the train-clean-100 hours training setup. The best configuration parameters discovered were 256 rows and ten columns.\ We also tried to run with a different MANN, the DNC \cite{graves2016hybrid}, which is the follow-up of the NTM system, but for the same chosen parameters, the NTM gave the best results in preliminary experiments.

At last, we used the KenLM toolkit \cite{heafield2011kenlm} with Kneser-Ney smoothing to train a 6-gram LM.\ For that, we used the Librispeech LM corpus from Kaldi and applied the BPE model to transform all words into sub-word units.\
The beam size is 60, and the weight for the 6-gram LM is 1. Furthermore, we report results with and without LM.

\subsection{Results}

Table \ref{tab:wer-cer-results} compares the performance of our proposed architecture, Conformer-NTM, versus the ESPnet conformer baseline, with and without an LM.\ 

Regarding the "Full Set" column results, without an LM, the Conformer-NTM slightly improves upon the Conformer baseline for test clean and test other in both training settings (train-clean-100 and train-960), except for the test other in the train-960 setting, where the Conformer-960h model achieves 6.7\%/2.9\% WER/CER and the Conformer-NTM-960h model achieves 6.8\%/2.8\% WER/CER.\

Decoding with the LM, our proposed models still achieve the lowest WERs and CERs compared to the baseline model, except for the train-960 setting, where the results in terms of WER are the same for the Conformer and Conformer-NTM.\ 


\subsection{Analysis}

To examine the behaviour of the Conformer-NTM model for longer sentences, we decided to create subsets from the test clean and the test other sets containing long and very long utterances as described in Section \ref{subsec:setup}, and evaluate its performance under these conditions.

\subsubsection{Long Utterances}


Regarding long utterances, without LM, we can observe from Table \ref{tab:wer-cer-results} ("Long - 100" column) that the Conformer-NTM model achieves the lowest WER and CER results compared to the baseline for both test clean and test other.\ For the train-clean-100 setting, the Conformer-NTM model improves from 7.5\% to 6.8\% in WER for the test clean and improves from 16.6\% to 15.8\% WER for the test other. For the train-960 setting, the Conformer-NTM model improves from 3.3\% to 3.0\% in WER for the test clean and from 5.3\% to 4.8\% WER for the test other.   

With LM, the Conformer-NTM model still obtains better results for both test clean and test other subsets, except for the train-960 setting in the test clean subset where the result of the Conformer is equal to the Conformer-NTM in WER.

\subsubsection{Very Long Utterances}

Concerning very long utterances, described in Section \ref{subsec:setup}, we can observe from Table \ref{tab:wer-cer-results} ("Very Long - 100" column) that the baseline conformer results start to degrade more when compared to the baseline conformer results in column "Long - 100", mainly because the distribution of lengths in concat-clean and concat-other is more distant from the distribution of lengths present in the training and development data.\ 

Additionally, with and without LM, the Conformer-NTM improves by a significant margin all the WER and CER scores when compared to the baseline conformer in both training settings, i.e., train-clean-100 and train-960. For instance, in the train-960 setting using LM, the Conformer-NTM achieves a relative WER reduction up to 58.1\% and 26.5\%  for the concat-clean and concat-other sets, respectively.    

These improvements demonstrate that the MANN based on the NTM memory helps the E2E Conformer ASR system to generalise better for very long sentences not seen during training without relying on any pre-processing of the data or changing training and decoding strategies when compared to traditional E2E ASR. Furthermore, the presence of the LM does not affect the improvements of the Conformer-NTM when compared to the conformer baseline. We hypothesise that the NTM memory learns to create more extended contexts at an acoustic level which benefits the decoder of the conformer when making the inference step.      

\section{Conclusions and Future Work}

In this work, we propose a new architecture, Conformer-NTM, that combines a MANN (based on the NTM) with a conformer for E2E ASR.\ We demonstrated that including the external memory is relevant to enhance the performance of the E2E ASR system for long utterances.\ Also, we observed that the Conformer-NTM starts to be more effective when the distribution length of the test data gets further away from the distribution length of the training data.\ Furthermore, in the presence of an LM and for very long utterances, the Conformer-NTM in the train-960 setting achieves a 58.1\% WER relative reduction for the concat-clean set and a 26.5\% WER relative reduction to the concat-other when compared to the Conformer baseline.   
Our future work includes investigating the effect of the NTM on other SOTA E2E ASR architectures. 

\bibliographystyle{IEEEtran}
\bibliography{template}

\begin{thebibliography}{10}
\providecommand{\url}[1]{#1}
\csname url@samestyle\endcsname
\providecommand{\newblock}{\relax}
\providecommand{\bibinfo}[2]{#2}
\providecommand{\BIBentrySTDinterwordspacing}{\spaceskip=0pt\relax}
\providecommand{\BIBentryALTinterwordstretchfactor}{4}
\providecommand{\BIBentryALTinterwordspacing}{\spaceskip=\fontdimen2\font plus
\BIBentryALTinterwordstretchfactor\fontdimen3\font minus
  \fontdimen4\font\relax}
\providecommand{\BIBforeignlanguage}[2]{{%
\expandafter\ifx\csname l@#1\endcsname\relax
\typeout{** WARNING: IEEEtran.bst: No hyphenation pattern has been}%
\typeout{** loaded for the language `#1'. Using the pattern for}%
\typeout{** the default language instead.}%
\else
\language=\csname l@#1\endcsname
\fi
#2}}
\providecommand{\BIBdecl}{\relax}
\BIBdecl

\bibitem{hmms}
M.~Gales and S.~Young, ``The application of hidden markov models in speech
  recognition,'' \emph{Foundations and Trends® in Signal Processing}, vol.~1,
  no.~3, pp. 195--304, 2008.

\bibitem{pmlr-v32-graves14}
A.~Graves and N.~Jaitly, ``Towards end-to-end speech recognition with recurrent
  neural networks,'' in \emph{Proceedings of the 31st International Conference
  on Machine Learning (ICML)}, ser. Proceedings of Machine Learning Research,
  vol.~32.\hskip 1em plus 0.5em minus 0.4em\relax PMLR, Jun 2014, pp.
  1764--1772.

\bibitem{8706675}
Z.~Xiao, Z.~Ou, W.~Chu, and H.~Lin, ``Hybrid ctc-attention based end-to-end
  speech recognition using subword units,'' in \emph{2018 11th International
  Symposium on Chinese Spoken Language Processing (ISCSLP)}.\hskip 1em plus
  0.5em minus 0.4em\relax IEEE, May 2018, pp. 146--150.

\bibitem{7472621}
W.~Chan, N.~Jaitly, Q.~Le, and O.~Vinyals, ``Listen, attend and spell: A neural
  network for large vocabulary conversational speech recognition,'' in
  \emph{2016 IEEE international conference on acoustics, speech and signal
  processing (ICASSP)}.\hskip 1em plus 0.5em minus 0.4em\relax IEEE, May 2016,
  pp. 4960--4964.

\bibitem{chiu2018state}
C.~Chiu, T.~Sainath, Y.~Wu, R.~Prabhavalkar, P.~Nguyen, Z.~Chen, A.~Kannan,
  R.~Weiss, K.~Rao, E.~Gonina \emph{et~al.}, ``State-of-the-art speech
  recognition with sequence-to-sequence models,'' in \emph{2018 IEEE
  International Conference on Acoustics, Speech and Signal Processing
  (ICASSP)}.\hskip 1em plus 0.5em minus 0.4em\relax IEEE, Apr 2018, pp.
  4774--4778.

\bibitem{graves2006connectionist}
A.~Graves, S.~Fern{\'a}ndez, F.~Gomez, and J.~Schmidhuber, ``Connectionist
  temporal classification: labelling unsegmented sequence data with recurrent
  neural networks,'' in \emph{Proceedings of the 23rd international conference
  on Machine learning (ICML)}, Jun 2006, pp. 369--376.

\bibitem{graves2012sequence}
A.~Graves, ``Sequence transduction with recurrent neural networks,'' \emph{In
  Proceedings of the 29th International Conference on Machine Learning (ICML)},
  2012.

\bibitem{vaswani2017attention}
A.~Vaswani, N.~Shazeer, N.~Parmar, J.~Uszkoreit, L.~Jones, A.~Gomez, L.~Kaiser,
  and I.~Polosukhin, ``Attention is all you need,'' in \emph{Advances in Neural
  Information Processing Systems (NIPS)}, vol.~30.\hskip 1em plus 0.5em minus
  0.4em\relax Curran Associates, Inc., 2017.

\bibitem{49414}
A.~Gulati, J.~Qin, C.~Chiu, N.~Parmar, Y.~Zhang, J.~Yu, W.~Han, S.~Wang,
  Z.~Zhang, Y.~Wu, and R.~Pang, ``{Conformer: Convolution-augmented Transformer
  for Speech Recognition},'' in \emph{Proc. Interspeech}, 2020, pp. 5036--5040.

\bibitem{narayanan2019recognizing}
A.~Narayanan, R.~Prabhavalkar, C.~Chiu, D.~Rybach, T.~Sainath, and T.~Strohman,
  ``Recognizing long-form speech using streaming end-to-end models,'' in
  \emph{IEEE Automatic Speech Recognition and Understanding Workshop (ASRU)},
  2019, pp. 920--927.

\bibitem{chiu2019comparison}
C.~Chiu, W.~Han, Y.~Zhang, R.~Pang, S.~Kishchenko, P.~Nguyen, A.~Narayanan,
  H.~Liao, S.~Zhang, A.~Kannan \emph{et~al.}, ``A comparison of end-to-end
  models for long-form speech recognition,'' in \emph{2019 IEEE Automatic
  Speech Recognition and Understanding Workshop (ASRU)}.\hskip 1em plus 0.5em
  minus 0.4em\relax IEEE, 2019, pp. 889--896.

\bibitem{hori2020transformer}
T.~Hori, N.~Moritz, C.~Hori, and J.~Le~Roux, ``Transformer-based long-context
  end-to-end speech recognition.'' in \emph{Proc. Interspeech}, October 2020,
  pp. 5011--5015.

\bibitem{hori2021conformer}
T.~Hori, N.~Moritz, C.~Hori, and J.~L. Roux, ``Advanced long-context end-to-end
  speech recognition using context-expanded transformers,'' in \emph{Proc.
  Interspeech}, August 2021, pp. 2097--2101.

\bibitem{bain2023whisperx}
M.~Bain, J.~Huh, T.~Han, and A.~Zisserman, ``Whisperx: Time-accurate speech
  transcription of long-form audio,'' \emph{arXiv preprint arXiv:2303.00747},
  2023.

\bibitem{51460}
W.~R. Huang, S.~yiin Chang, D.~Rybach, T.~N. Sainath, R.~Prabhavalkar,
  C.~Allauzen, C.~Peyser, and Z.~Lu, ``E2e segmenter: Joint segmenting and
  decoding for long-form asr,'' in \emph{Proc. Interspeech}, September 2022.

\bibitem{10045036}
B.~Lyu, C.~Fan, Y.~Ming, P.~Zhao, and N.~Hu, ``En-hacn: Enhancing hybrid
  architecture with fast attention and capsule network for end-to-end speech
  recognition,'' \emph{IEEE/ACM Transactions on Audio, Speech, and Language
  Processing}, vol.~31, pp. 1050--1062, 2023.

\bibitem{graves2014neural}
A.~Graves, G.~Wayne, and I.~Danihelka, ``Neural turing machines,'' \emph{arXiv
  preprint arXiv:1410.5401}, 2014.

\bibitem{sari2020unsupervised}
L.~Sar{\i}, N.~Moritz, T.~Hori, and J.~Roux, ``Unsupervised speaker adaptation
  using attention-based speaker memory for end-to-end asr,'' in \emph{IEEE
  International Conference on Acoustics, Speech and Signal Processing
  (ICASSP)}.\hskip 1em plus 0.5em minus 0.4em\relax IEEE, 2020, pp. 7384--7388.

\bibitem{dehak2010front}
N.~Dehak, P.~Kenny, R.~Dehak, P.~Dumouchel, and P.~Ouellet, ``Front-end factor
  analysis for speaker verification,'' \emph{IEEE Transactions on Audio,
  Speech, and Language Processing}, vol.~19, no.~4, pp. 788--798, 2010.

\bibitem{panayotov2015librispeech}
V.~Panayotov, G.~Chen, D.~Povey, and S.~Khudanpur, ``Librispeech: an asr corpus
  based on public domain audio books,'' in \emph{IEEE international conference
  on acoustics, speech and signal processing (ICASSP)}.\hskip 1em plus 0.5em
  minus 0.4em\relax IEEE, 2015, pp. 5206--5210.

\bibitem{wu2020streaming}
C.Wu, Y.Wang, Y.Shi, C.Yeh, and F.Zhang, ``{Streaming Transformer-Based
  Acoustic Models Using Self-Attention with Augmented Memory},'' in \emph{Proc.
  Interspeech 2020}, 2020, pp. 2132--2136.

\bibitem{tsunoo2019transformer}
E.Tsunoo, Y.Kashiwagi, T.Kumakura, and S.Watanabe, ``Transformer asr with
  contextual block processing,'' in \emph{IEEE Automatic Speech Recognition and
  Understanding Workshop (ASRU)}.\hskip 1em plus 0.5em minus 0.4em\relax IEEE,
  2019, pp. 427--433.

\bibitem{graves2016hybrid}
A.~Graves, G.~Wayne, M.~Reynolds, T.~Harley, I.~Danihelka,
  A.~Grabska-Barwi{\'n}ska, S.~Colmenarejo, E.~Grefenstette, T.~Ramalho,
  J.~Agapiou \emph{et~al.}, ``Hybrid computing using a neural network with
  dynamic external memory,'' \emph{Nature}, vol. 538, no. 7626, pp. 471--476,
  2016.

\bibitem{pmlr-v48-santoro16}
A.~Santoro, S.~Bartunov, M.~Botvinick, D.~Wierstra, and T.~Lillicrap,
  ``Meta-learning with memory-augmented neural networks,'' in \emph{Proceedings
  of The 33rd International Conference on Machine Learning}, ser. Proceedings
  of Machine Learning Research, vol.~48.\hskip 1em plus 0.5em minus 0.4em\relax
  PMLR, 20--22 Jun 2016, pp. 1842--1850.

\bibitem{kim2017joint}
S.~Kim, T.~Hori, and S.~Watanabe, ``Joint ctc-attention based end-to-end speech
  recognition using multi-task learning,'' in \emph{2017 IEEE international
  conference on acoustics, speech and signal processing (ICASSP)}.\hskip 1em
  plus 0.5em minus 0.4em\relax IEEE, 2017, pp. 4835--4839.

\bibitem{Povey_ASRU2011}
D.~Povey, A.~Ghoshal, G.~Boulianne, L.~Burget, O.~Glembek, N.~Goel,
  M.~Hannemann, P.~Motlicek, Y.~Qian, P.~Schwarz, J.~Silovsky, G.~Stemmer, and
  K.~Vesely, ``The kaldi speech recognition toolkit,'' in \emph{IEEE 2011
  Workshop on Automatic Speech Recognition and Understanding}, Dec. 2011.

\bibitem{watanabe2018espnet}
S.~Watanabe, T.~Hori, S.~Karita, T.~Hayashi, J.~Nishitoba, Y.~Unno, N.~Soplin,
  J.~Heymann, M.~Wiesner, N.~Chen \emph{et~al.}, ``Espnet: End-to-end speech
  processing toolkit,'' \emph{in Proc. ISCA Interspeech}, Sep 2018.

\bibitem{Park_2019}
D.~S. Park, W.~Chan, Y.~Zhang, C.~Chiu, B.~Zoph, E.~D. Cubuk, and Q.~Le,
  ``{SpecAugment}: A simple data augmentation method for automatic speech
  recognition,'' in \emph{Interspeech 2019}.\hskip 1em plus 0.5em minus
  0.4em\relax ISCA, Sep 2019.

\bibitem{sennrich-etal-2016-neural}
R.~Sennrich, B.~Haddow, and A.~Birch, ``Neural machine translation of rare
  words with subword units,'' in \emph{Proceedings of the 54th Annual Meeting
  of the Association for Computational Linguistics}.\hskip 1em plus 0.5em minus
  0.4em\relax Association for Computational Linguistics, Aug 2016.

\bibitem{heafield2011kenlm}
K.~Heafield, ``Kenlm: Faster and smaller language model queries,'' in
  \emph{Proceedings of the sixth workshop on statistical machine translation},
  2011, pp. 187--197.

\end{thebibliography}

\end{document}